\documentclass[a4paper]{article}

\usepackage{17lomcon}        % Proceedings volume layout metrics
\usepackage{cite}             % Smart range citations
\usepackage{epsfig}           % Encapsulated PostScript figure inclusion

\begin{document}

%%%%%%%%%%%%%%%%%%%%%%%%%%%%%%%%%%%%%%%%%%%%%%%%%%%%%%%%%%%%%%%%%%
% The preamble of the paper
%%%%%%%%%%%%%%%%%%%%%%%%%%%%%%%%%%%%%%%%%%%%%%%%%%%%%%%%%%%%%%%%%%

\title{EXOTIC HIGGS DECAYS IN U(1) EXTENSIONS OF THE MSSM}

\author{Peter Athron}

\affiliation{ARC Centre of Excellence for Particle Physics at the Terascale,
School of Physics, Monash University, Melbourne, Victoria 3800, Australia}

\author{Margarete M\"{u}hlleitner}

\affiliation{Institute for Theoretical Physics, Karlsruhe Institute of Technology,
76128 Karlsruhe, Germany}

\author{Roman Nevzorov and Anthony G. Williams}

\affiliation{ARC Centre of Excellence for Particle Physics at the Terascale and CSSM,
School of Chemistry and Physics, University of Adelaide, Adelaide SA 5005, Australia}

% You may repeat \author and \affiliation as many times as necessary!

\date{}
% Print it out!
\maketitle

%%%%%%%%%%%%%%%%%%%%%%%%%%%%%%%%%%%%%%%%%%%%%%%%%%%%%%%%%%%%%%%%%%
% The preamble of the paper
%%%%%%%%%%%%%%%%%%%%%%%%%%%%%%%%%%%%%%%%%%%%%%%%%%%%%%%%%%%%%%%%%%

\begin{abstract}
We study the decays of the lightest CP--even Higgs boson into a pair of pseudoscalar Higgs states within $U(1)_N$ extensions of the MSSM.
\end{abstract}

\section{Introduction}

A new bosonic state with a mass around $125\,\mbox{GeV}$,
which was discovered at the LHC a few years ago, can provide a window into
new physics beyond the Standard Model (SM). The
signal strengths associated with this particle are currently consistent with
the SM Higgs boson. However further exploration of this resonance
can lead to new exciting discoveries. Physics beyond the SM may change
the Higgs decay rates to SM particles and give rise to new decay modes.
Here we consider the decays of the SM-like Higgs boson into a pair of
pseudoscalar states within $U(1)$ extensions of the Minimal Supersymmetric
Standard Model (MSSM). In particular we focus on the $E_6$ inspired
SUSY models which are based on the low--energy SM gauge group together
with an extra $U(1)_{N}$ gauge symmetry defined by:
\begin{equation}
U(1)_N=\frac{1}{4} U(1)_{\chi}+\frac{\sqrt{15}}{4} U(1)_{\psi}\,.
\label{1}
\end{equation}
Gauge $U(1)_{\psi}$\quad and $U(1)_{\chi}$ symmetries can
originate from the breakings $E_6\to$ $SO(10)\times U(1)_{\psi}$,
$SO(10)\to SU(5)\times U(1)_{\chi}$. Different aspects of the phenomenology
of these $E_6$ inspired supersymmetric (SUSY) models have been extensively studied in
\cite{King:2005jy,King:2005my,King:2007uj,King:2008qb,Athron:2009ue,Athron:2009bs,Hall:2010ix,Athron:2011wu,Nevzorov:2012hs,Athron:2012sq,Nevzorov:2013tta,Nevzorov:2013ixa,Athron:2014pua}.

\section{The $U(1)_N$ extensions of the MSSM}

These SUSY models imply that just below the scale $M_X$ the gauge symmetry
in the $E_6$ Grand Unified Theory (GUT) is broken down to
$SU(3)_C\times SU(2)_W\times U(1)_Y\times U(1)_{N}\times Z_{2}^{M}$,
where $Z_{2}^{M}=(-1)^{3(B-L)}$ is a matter parity. To ensure anomaly
cancellation, gauge coupling unification and the appropriate breakdown of
the gauge symmetry at low energies the matter content of the SUSY models
under consideration involves three $27$ representations of the $E_6$,
two $SU(2)_W$ doublets $L_4$ and $\overline{L}_4$ from extra $27'_1$
and $\overline{27'}_1$, two SM singlets $S$ and $\overline{S}$ that carry
non--zero $U(1)_N$ charges and stem from another pair of $27'_2$
and $\overline{27'}_2$ as well as a SM singlet $\phi$ that does not participate
in gauge interactions. Thus these models involve extra matter beyond the
MSSM. In particular, they include three pairs of $SU(2)_W$--doublets
($H^d_{i}$ and $H^u_{i}$) that have the quantum numbers of Higgs doublets,
and three pairs of colour triplets of exotic quarks $\overline{D}_i$ and $D_i$ with
electric charges $+ 1/3$ and $-1/3$ respectively. To suppress flavour changing
neutral currents as well as the most dangerous baryon and lepton number violating
operators one can impose a discrete $\tilde{Z}^{H}_2$ symmetry, under which
all superfields except one pair of $H^{d}_{i}$ and $H^{u}_{i}$ (say
$H_d\equiv H^{d}_{3}$ and $H_u\equiv H^{u}_{3}$), $L_4$, $\overline{L}_4$,
$S$, $\overline{S}$ and $\phi$ are odd.

The conservation of $Z_{2}^{M}$ and
$\tilde{Z}^{H}_2$ symmetries implies that these models contain at least two
states which are absolutely stable and can contribute to the dark matter density.
One of these states is a lightest SUSY particle (LSP) whereas another one is
the lightest ordinary neutralino. Using the method proposed in \cite{Hesselbach:2007te}
it was argued that there is a theoretical upper bound on the mass of the LSP
\cite{Hall:2010ix}. In the simplest phenomenologically viable scenarios the LSP
has a mass around $1\,\mbox{eV}$ or even smaller forming hot dark matter
in the Universe. The presence of such very light neutral fermions in the particle spectrum
may have interesting implications for neutrino physics (see, for example \cite{Frere:1996gb}).
In these scenarios the lightest ordinary neutralino can account for all or some of the observed
dark matter density.

\section{Non-Standard Higgs Decays}

The Higgs sector of the SUSY models under consideration contains two Higgs
doublets $H_d$ and $H_u$ as well as $S$, $\overline{S}$ and $\phi$.
The scalar components of the superfields $S$ and $\overline{S}$ acquire
large vacuum expectation values (VEVs) that break the $U(1)_N$ gauge symmetry inducing
the masses of exotic particles and the $Z'$ boson. The interaction of $S$, $\overline{S}$
and $\phi$ is determined by the following terms in the superpotential that can be written as
\begin{equation}
W_{S}= \sigma \phi S \overline{S} + \frac{\kappa}{3}\phi^3+\frac{\mu}{2}\phi^2+\Lambda\phi\,.
\label{2}
\end{equation}
When~ the~ couplings~ $\kappa$,~ $\mu$ and $\Lambda$ are rather small these models possess~
an~ approximate global $U(1)$ symmetry. This symmetry is spontaneously broken by the
VEVs of $S$, $\overline{S}$ and $\phi$ leading to the presence of a light pseudoscalar
state in the particle spectrum. Our analysis indicates that the couplings of this state to the
SM particles are always small. Therefore even if this pseudoscalar is substantially lighter
than $100\,\mbox{GeV}$ it could escape detection at former and present collider experiments.
Moreover this state can be so light that it gives rise to the decays of the SM-like Higgs boson
into a pair of these pseudoscalar states. The branching ratio of such decays depends
rather strongly on the values of the couplings that explicitly violate the global $U(1)$ symmetry.

To simplify our analysis we set $\mu=0$ as well as $\Lambda=0$ and examine the dependence
of the branching ratio of the exotic Higgs decays on $\kappa$. For $\kappa\sim 0.001$ the pseudoscalar
Higgs boson with a mass of $40-60\,\mbox{GeV}$ can be obtained without fine-tuning but
the branching ratio of the SM-like Higgs decays into a pair of such particles is smaller
than $10^{-4}$ \cite{Athron:2014pua}. Thus it will be rather difficult to observe such decays at the LHC.
If $\kappa > 0.01$ a fine tuning of at least $1\%$ is required to get the lightest
CP-odd Higgs state with mass of $40-60\,\mbox{GeV}$. However the branching ratio of the
corresponding exotic Higgs decays can be larger than $1\%$ in this case \cite{Athron:2014pua}.
After being produced from such decays the lightest pseudoscalars decay into a pair of
$b$-quarks or $\tau$-leptons resulting in four fermion final states.

\section*{Acknowledgments}

This work was supported by the University of Adelaide and the
Australian Research Council through the ARC Center of Excellence in
Particle Physics at the Terascale.

%%%%%%%%%%%%%%%%%%%%%%%%%%%%%%%%%%%%%%%%%%%%%%%%%%%%%%%%%%%%%%%%%%
% References
%%%%%%%%%%%%%%%%%%%%%%%%%%%%%%%%%%%%%%%%%%%%%%%%%%%%%%%%%%%%%%%%%%

\end{document}